\documentclass[showpacs,preprint,prl]{revtex4}

\usepackage{graphicx}
\usepackage{subfigure}

\newcommand{\figurewidth}{0.42\textwidth}

\begin{document}

\title{Stick-release pattern in stretching single condensed polyelectrolyte toroids}
\author{Paul C\'ardenas-Lizana}
\author{Pai-Yi Hsiao}
\email[Corresponding author, E-mail: ]{pyhsiao@ess.nthu.edu.tw}
\affiliation{Department of Engineering and System Science, National Tsing Hua University, 
Hsinchu, Taiwan 300, R.O.C.}
\date{\today} 

\begin{abstract}
Using Langevin dynamics simulations, we study elastic response of single 
semiflexible polyelectrolytes to an external force pulling on the chain ends,
to mimic the stretching of DNA molecules by optical tweezers. 
The linear chains are condensed by multivalent counterions into toroids.
The force-extension curve shows a series of sawtooth-like structure, known as 
the stick-release patterns in experiments. 
We demonstrate that these patterns are a consequence of the loop-by-loop unfolding 
of the toroidal structure. 
Moreover, the dynamics, how the internal structure of chain varies under tension,  
is examined.  At the first stage of the stretching, the toroidal condensate decreases 
its size until the loss of the first loop in the toroid and then, 
oscillates around this size for the rest of the unfolding process. 
The normal vector of the toroid is pulled toward the pulling-force   
direction and swings back to its early direction repeatedly when the toroidal chain 
looses a loop. 
The results provide new and valuable information concerning the elasticity and the 
microscopic structure and dynamic pathway of salt-condensed DNA molecules being stretched.
\end{abstract}

\pacs{82.35.Rs, 87.15.La, 87.15.hp, 87.15.ap}

\maketitle

There are continually strong demands in understanding the properties of DNA 
molecules because of their broad implications in biology and the benefits in gene 
therapy~\cite{Vijayanathan02}.
Owing to the progress of nano-technology, researchers are now allowed to investigate 
the elastic response of DNA under the action of an external force, 
at the level of single molecules~\cite{Smith92}. 
When a DNA molecule was stretched, different kinds of force-extension curve 
(FEC) have been observed~\cite{Baumann00,Murayama01,Murayama03}, 
depending on the multivalent counterion concentration, $C$, in the solution. 
For $C$ staying outside the region bounded by the condensation threshold $C_c$ and 
the decondensation threshold $C_d$ ($C_c<C_d$), DNA molecules are in a coil state 
and FEC is basically described by the Marko-Siggia formula derived 
from the wormlike chain (WLC) model~\cite{Marko95}. 
For $C$ inside the region, 
the DNAs are collapsed into globule of ordered structure and the preferable structure is toroid~\cite{Bloomfield96}.
At this moment, FEC shows \textit{plateau force} while $C \sim C_c$, 
or \textit{stick-release pattern} while $C$ is close to $C_0$, where $C_0$ is 
the salt concentration at which the effective chain charge is neutralized by 
the multivalent counterions~\cite{Nguyen00}. 
For the plateau force, the elastic response is constant over a wide range of 
chain extension and its value is larger than the prediction of the WLC model,
whereas for the stick-release pattern, the force is periodic and piecewisely 
described by the Marko-Siggia formula~\cite{Baumann00,Murayama03}.

It was suggested that the plateau force is derived from the fact that
the ratio of the extension to the effective chain contour length is a constant
and thus produces a constant force according to the Marko-Siggia formula~\cite{Wada02}. 
To explain the stick-release pattern, two models have been proposed. 
The first model~\cite{Wada02,Ueda96} is based upon the observation of 
the \textit{rings-on-a-string} structure of a long DNA chain. 
The formation of this gripping pattern could be a consequence 
of the abrupt breaking of one ring into several under the stretching 
of an external force.
Elastic response between any two ring breakings is determined 
by the coil part of the chain and follows the WLC model. 
The second model~\cite{Wada05,Murayama03}, on the other hand, attributes 
this phenomenon to a result of the loop-by-loop unfolding of a toroidal DNA 
condensate under stretching, because a pronounced quantization in the 
DNA release length of about 300nm has been demonstrated, which correlates
exactly with the periphery length of a typical  DNA toroid of size 
$R\simeq 50$nm. 
Nevertheless, a concrete evidence is still missing and 
researchers cannot make a conclusive connection of the stick-release pattern 
with the second model yet.

Recently, the pathway of unwrapping a spool of DNA chain helically coiled onto 
a histone protein has been studied by theorists to investigate the stability and dynamics
of nucleosomes under tension~\cite{Kulic04}. 
They pointed out the similarity between two seemingly unrelated problems:
unwrapping of nucleosomes and unfolding of DNA toroidal condensates,
and predicted a catastrophic event for the two systems under tension:   
a sudden and quantized unraveling happened once a time for a DNA turn.
This unwrapping pathway is difficult to be investigated by experiments and
the information concerning the internal structure of toroidal DNA condensates
under tension is still not complete. 
Therefore, in this study, we conduct computer simulations to study 
the elastic response of a toroidal polyelectrolyte (PE) condensate and 
investigate structural variation and unfolding dynamics of chain 
under tension. 
We focus on the case that the amount of multivalent counterions 
is in charge equivalence with the PE chain.
This choice gives us the most chance to observe the stick-release pattern.
There have been simulation works devoting to the study of PE chains under 
tension~\cite{Wada05,Marenduzzo04,Lee04}. 
However, in most of the works, the chains were flexible and condensed 
into disordered structures either by compacting potentials 
or by monovalent counterions with a ultra-strong Coulomb coupling. 
Our work here, on the contrary, deals with a more realistic situation 
in an ambient condition where the chain is semiflexible, collapsed by 
multivalent counterions into a toroid. 
This kind of study is still few~\cite{Khan05}, and involves nonlinear 
and nonequilibrium effects coming from the arrangement of ions, 
the long-range Coulomb interaction, and the pulling speed.

Our system contains a single chain and many counterions. 
The chain is constituted of $N_m=512$ monomers, each of which
carries a negative unit charge $-e$. 
The counterions are tetravalent and the number of the counterions is 128.
The excluded volume of the monomers and the counterions is modeled by 
the Lennard-Jones potential $\varepsilon_{\rm LJ}\left[2(\sigma/r)^6-1\right]^2$
truncated at the minimum. 
We assumed an identical $\varepsilon_{\rm LJ}$ for the monomers and counterions 
but different diameter $\sigma$. 
The diameter of the counterions $\sigma_{c}$ is half of that of the monomers 
$\sigma_{m}$.
In the following text, we use $\varepsilon_{\rm LJ}$ and $\sigma_{c}$
as the units of energy and length, respectively.  
Two consecutive monomers on the chain are connected by a bond of length $b$ 
via the harmonic potential $k_b(b-b_0)^2$ with $k_b=100$ and $b_0=1.1$. 
The chain stiffness is described by an angle potential 
$k_2(\theta - \theta_0)^2+ k_4 (\theta - \theta_0)^4$ with $k_2=5$, 
$k_4=200$ and $\theta_0=\pi$, where $\theta$ is the angle between two 
adjacent bonds on the chain.  
Coulomb interaction is expressed by $\lambda_{\rm B}k_BT z_i z_j/r$ 
where $r$ is the separation distance of two particles of valences $z_i$ and 
$z_j$, and $\lambda_{\rm B}$ is the Bjerrum length. 
We set $\lambda_{\rm B}=4.68$ and the temperature $k_BT=1.2$. 
We assumed further that the bond potential and the angle potential have already 
incorporated the effect of non-bonded interactions along the chain, 
including the excluded volume interaction and Coulomb interaction.
Therefore, the non-bonded interactions are deactivated 
for pairs of monomers separated by less than three bonds on the chain.
This kind of setup has been used in simulations.
The advantage of this setup is that the bond and the angle potentials given here 
have been 
the \textit{entire} potentials without need to take into account further the
non-bonded interactions between the monomers. 
Please notice that the deactivation of Coulomb interaction locally on the chain 
will not violate the requirement of charge neutrality because it is a global 
property of a system and the total charge is always zero.
The system is placed in a rectangular box of size
$620 \times 91 \times 91$  and periodic boundary condition is applied.
The technique of PPPM Ewald sum is employed to calculate the 
Coulomb interaction.  
We performed Langevin dynamics simulations~\cite{note-tools}  
with the friction coefficient $\zeta$ setting to $2\tau^{-1}$ and $1\tau^{-1}$ for 
the monomers and the counterions, respectively, where 
$\tau=\sigma_c\sqrt{m/\varepsilon_{\rm LJ}}$ is the time unit and
$m$ is the particle mass.
Stevens has used a similar model~\cite{Stevens01} 
and shown that a PE chain can collapse into a toroid 
due to the condensation of tetravalent counterions.
Since the line charge density matches that of a double-stranded DNA (dsDNA)
and the size of monomer matches the size of a phosphate group, our model can 
be used to understand the system of DNA condensed into a toroid.
We remark that the persistence length in our model is one order of magnitude shorter 
than dsDNA. The reason for this setup is to allow the formation of a toroid in a short
chain length; thus the simulations become feasible under limited computing 
resources.

\begin{figure}
\begin{center}
\includegraphics[angle=270,width=\figurewidth]{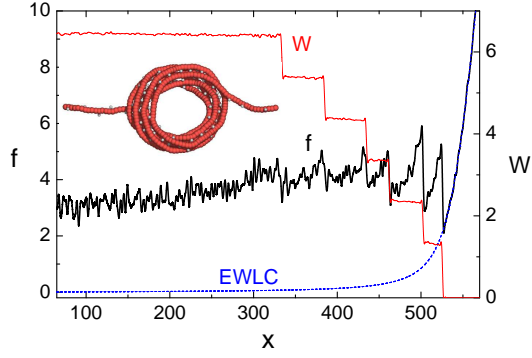} 
\caption{Force $f$ vs.~extension $x$ (thick solid curve). 
The dashed line is the fitting curve for $x \ge 526$ by the EWLC model. 
$W$ is plotted in solid curve and the value is read from the right axis of 
ordinate. The inset is a snapshot of the chain before stretching. }
\label{fig:FEC}
\end{center}
\end{figure}

The initial configuration of chain is a circular helix with its central line 
lying on the $x$ axis. This configuration is used to favor the formation of a toroid. 
Other structures, such as cigar or racket shapes, sometimes appear 
in the simulations~\cite{Stevens01,Wei07}. All these structures have been 
observed in experiments of DNA condensation and people believe that toroid is the 
most stable structure~\cite{Bloomfield96}. 
Since different condensed structure could produce different FEC under tension,  
in this study we focus on the case of chains collapsing into toroids. 
The two chain ends are guided by a spring force towards two points 
on the $x$-axis with separation distance equal to 60. After reaching a stable state, 
most of the chains form \textit{rod-toroid-rod} structure as shown in the inset of Fig.~\ref{fig:FEC}. 
Since chain ends are constrained, knots will not appear on the chain.
We have chosen different radii and pitches of helix as our initial configurations and 
verified that the size of the generated toroid is independent of these choices.
The toroid size is defined by two radii, the minor radius $r_0$ and the major radius $R_0$.
The previous one is the radius of the annular tube of the toroid and  
the latter is the radius measured from the toroid center to the center of the annular tube.
These two radii determine the gyration tensor of a toroid with  
the three eigenvalues equal to  $r_0^2/4$, $(4R_0^2+3r_0^2)/8$, and $(4R_0^2+3r_0^2)/8$. 
The gyration tensor of a set of $N$ particles can be calculated by simulations 
according to the equation 
$T_{\alpha\beta}=\sum_{i=1}^{N}(\vec{r}_i-\vec{r}_{cm})_{\alpha}(\vec{r}_i-\vec{r}_{cm})_{\beta}/N$ 
where $\vec{r}_{cm}$ is the center of mass of the set of particles
and the subscripts $\alpha$ and $\beta$ denote the three Cartesian components.
We calculated the gyration tensor of the toroid on the chain and estimated  
$r_0$ and $R_0$ from the three eigenvalues $\lambda_i$ ($i=1,2,3$).
The results are $r_0=3.60(1)$ and $R_{0}=12.3(1)$.
It is known that a ring structure ($r_0\ll R_0$) can be characterized by a quantity,
called \textit{asphericity}, defined as 
$A=\sum_{i=1}^{3}\sum_{j=1}^{3}(\lambda_i-\lambda_j)^2/(2\sum_{i=1}^{3}\lambda_i)^2$,
with its value equal to $0.25$.
Since our toroid  has a finite $r_0$, $A$ is smaller than 0.25 and takes a value of
$0.235(1)$.
The winding number $W$ is another important quantity to describe the status of a toroid, 
which counts the number of turns winding around the toroid central axis. 
Our toroidal condensates have $W=6.5$ before stretching. 
The decimal $0.5$ in $W$ reflects the fact that the two chain ends stay on the opposite sides 
of the toroid. 

We fixed one end of the condensed PE chains and pulled the other end outwards
along the $x$-axis using a spring force with constant pulling speed $v=0.0005$.
Since unfolding of a condensed PE is a nonequilibrium process, it is delicate
to choose the pulling speed~\cite{Lee04,Wada05}.  We have verified that
significant nonequilibrium effect is detected only when the pulling speed
excesses the characteristic speed, $v_s=R_0/\tau_R\sim 0.002$, estimated by the
Rouse relaxation time $\tau_R$.  Our choice of $v$ is slow enough to minimize
the dependence of FEC on the pulling speed, which essentially probes the
elastic response of PEs in the limit of zero pulling speed~\cite{Lee04}.
Nevertheless, it does not mean that our chains reach equilibrium for every
stretched distance. To study it, one needs to perform equilibrium simulations
with the chain ends fixed on every stretched distance. However, this sort of approach
demands much more computing resources.  For practical purpose to understand the
stick-release pattern of a toroidal DNA condensate under tension, we decided to
follow the arguments given by Lee and Thirumalai~\cite{Lee04} that the
nonequilibrium method with a small pulling speed can be a good approach,
sufficient enough to study the problems of the elasticity, the internal
structure, and the dynamics.

We prepared 20 independent toroids and performed stretching process. 
A typical FEC obtained in our simulations is shown in Fig.~\ref{fig:FEC}.
The extension $x$ is the distance between two chain ends.
The data have been passed through a low-pass filter to attenuate the 
high-frequency noise. 
We have verified that the filter does not modify the content of the force curve.  
The FEC shows approximately a plateau force up to  $x\approx270$. 
After that, the force increases and then decreases abruptly with 
extension. 
This tooth-like structure repeatedly appears in the FEC 
and becomes shaper and sharper as the extension increases. 
This is so-called stick-release pattern, which has been  
experimentally reported~\cite{Baumann00,Murayama03}. 
We plot in the same figure the winding number $W$. 
We see that $W$ is a downstairs function and 
the height of each stairstep is equal to one.
It shows that the toroidal PE looses one loop, followed by one loop, under stretching. 
The results support the picture of Murayama \textit{et al.}~\cite{Murayama03}. 
The dynamics that one loop breaks into several was not observed in this study. 
Moreover, we observed that the sawteeth in FEC appear coherently 
with the steps of $W$. 
Each sawtooth corresponds exactly to one step in $W$. 
It demonstrates that the stick-release pattern is a consequence of the loop-by-loop 
unfolding of the condensed PE toroid, and confirms the theoretical picture of 
stepwise unwinding of PE under stretching~\cite{Tamashiro00}. 

In the final stage of the stretching ($x>525$), $f$ increases largely 
and the elastic response is similar to the WLC model. 
Since our chain is extensible, we fitted these data by the modified Marko-Siggia 
formula~\cite{Odijk95} derived from the extensible WLC (EWLC) model, 
\begin{equation}
\frac{fP}{k_{B}T}=\frac{x}{L}+\frac{1}{4\left(1-x/L+f/K\right)^2}-\frac{1}{4}-\frac{f}{K}
\label{eq:EWLC}
\end{equation}
where $L$ is the contour length, $P$ is the persistence length, and 
$K$ is the elastic modulus. 
We obtained $L=559.7$, $P=25.2$ and $K=230.2$ by fitting in the region $x>525$.
The fitting curve $f_{\rm EWLC}$, plotted in dashed line  in Fig.~\ref{fig:FEC}, 
matches well with the FEC in this region. 
The fitting values are consistent with the theoretical ones, 
$L=(N_m-1)b_0=562.1$ and $K=2k_b b_0=220$.
It is known that $P$ is a sum of two contributions; 
one is the intrinsic persistence length $P_0$ originating 
from the chain bare stiffness, and the other is the electrostatic
persistence length $P_e$ owing to the Coulomb repulsion 
between monomers on the chain. 
In this study, we calculated $P_0$  by equating the bending energy 
$E_{\rm bend}$ to $k_BT/2$ for a chain segment of length $P_0$ 
with the bending angle equal to 1 rad. 
$E_{\rm bend}$ is equal to $\left(k_2(b_0/P_0)^2+k_4(b_0/P_0)^4\right)\times(P_0/b_0)$
because there are $P_0/b$ monomers on the chain segment and 
the bond angle $\theta$ at each monomer is $\pi-(b_0/P_0)$ in average. 
By solving this equation, we have $P_0=12.2$,
which gives an estimation of $P_e$ equal to $P-P_0=13.0$. 
It is worth noticing that $P_0$ obtained here is approximately  equal to 
the major radius $R_0$ of the toroidal condensate before stretching. 
This is a result of charge neutralization.
The condensed tetravalent counterions neutralize the chain charge
and therefore, the electrostatics does not play a major  role 
in determination of the toroid size. 
At this moment, $P_e$ is roughly zero as shown 
in the previous study of flexible chains~\cite{Hsiao06b}.     
So the major radius relates directly the intrinsic 
persistence length of a PE.  

We calculated the excess work for the toroidal condensate 
$\Delta W=\int_{60}^{525} (f-f_{\rm EWLC})dx$, and found 
$\Delta W=1658 $. Therefore, the condensation energy 
is $3.47$ per monomer (because the toroid part of the chain contains 478 monomers before stretching).
This energy is mainly determined by the energy needed to break 
an \textit{electrostatic binding} between a condensed tetravalent counterion 
and a monomer, which reads as $|\lambda_B k_BT (+4)(-1)/1.5|=14.98$
or equivalently $3.74$ per monomer.
This finding suggests an electrostatic origin for the PE chain condensation.

\begin{figure}
\begin{center}
\includegraphics[angle=270,width=\figurewidth]{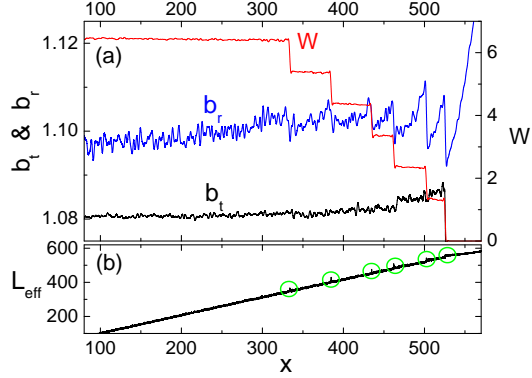} 
\caption{(a) $b_t$ and $b_r$, (b) $L_{\rm eff}$ as a function of $x$.
$W$ is plotted in the figure (a) to indicate the occurrence of 
the loop-by-loop unfolding.}
\label{fig:BL}
\end{center}
\end{figure}

The condensed chain preserves the rod-toroid-rod structure in most of the time 
during stretching.  
In order to understand more deeply the elastic response, 
we calculated the average bond length on the toroid part, $b_t$,
and on the rod part, $b_r$, of the chain. 
The results are presented in Fig.~\ref{fig:BL}(a). 
We found that $b_t$ stays roughly a constant up to $x\simeq 450$, and
the value is 1.08. This value is smaller than the equilibrium 
value $b_0=1.1$ of a bond, obviously due to the condensation of 
the tetravalent counterions which tightly bind middle part of the chain 
to form a toroidal structure. It is this tight binding 
which makes shorter the bond length $b_t$.  
Since the toroid part consists of several loops,  
the effective elastic modulus on this part is very stiff.
Therefore, the variation of $b_t$ with the extension $x$ is hardly seen. 
On the other hand, $b_r$ on the rod part displays a sawtooth structure,
coinciding with the FEC.
This shows that the elastic response of the chain is mainly determined by
the rod part.
If $f$ is strong enough to overcome the binding between 
the strands on the toroid, a loop is pulled out of it.

We also calculated the effective contour length of the chain $L_{\rm eff}$, 
defined as the total length on the rod part plus the diameter of the 
toroid. The result is shown in Fig.~\ref{fig:BL}(b).
We see that $L_{\rm eff}$, on the main trend, depends linearly on $x$, 
but, locally, shows a series of small peaks 
(enclosed inside circles in the figure). 
Each peak corresponds to the moment when a loop is pulled out of the toroid and 
the tension of the chain is suddenly released; 
consequently, $f$ is sharply decreased and so is $b_r$. 
The linear dependence of $L_{\rm eff}$ tells us that $x/L_{\rm eff}$ is approximately 
constant during the pulling process, which suggests the existence of a reference force $f_0$. 
$f_0$ can be estimated by replacing $x/L$ in Eq.~(\ref{eq:EWLC}) by $x/L_{\rm eff}$. 
Since $x/L_{\rm eff}\simeq 0.95$ in this study,
$f_0$ is about 3.5.  

We did find that the sawtooth structure of FEC oscillates around this reference force.
These oscillations can be attributed to the fluctuations of $L_{\rm eff}$ (or $b_r$) due to the loop-by-loop 
unfolding of the chain.
This phenomenon has been observed experimentally~\cite{Baumann00,Murayama01,Murayama03}.  
The observed reference force can be used to estimate $x/L_{\rm eff}$ in experiments 
by $x/L_{\rm eff}=1-\sqrt{k_BT/4Pf_0}$~\cite{Marko95}, 
which gives a typical value of $0.8$ for DNA condensation, for e.g., by spermidine. 
This value is smaller than our simulation value 0.95. 
The discrepancy is  due to the higher counterion valency 
and the smaller ion size used in this study, which leads to a 
stronger binding force between toroid loops than in the real experiments
 and hence, a larger $f_0$.  Consequently, the condensation energy 
obtained here is higher than a typical value obtained by experiments,
$0.08$ to $0.3 k_BT$ per base pair~\cite{Baumann00,Murayama01,Murayama03}. 

\begin{figure}
\begin{center}
\includegraphics[angle=270,width=\figurewidth]{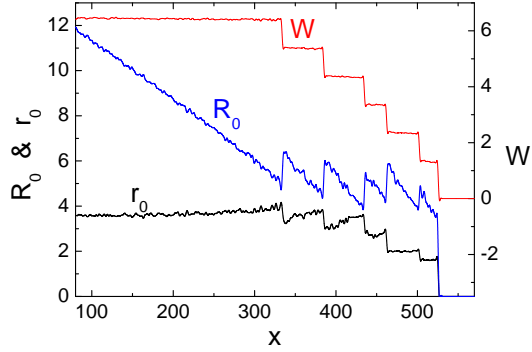} 
\caption{$R_0$, $r_0$ and $W$ as a function of $x$.}
\label{fig:R0_r0}
\end{center}
\end{figure}

We further investigated how the size of the toroid varies under stretching
and the results are shown in Fig.~\ref{fig:R0_r0}.
We observed that the major radius $R_0$ decreases linearly until 
the release of the first loop. 
It abruptly increases at the moment when a loop is pulled out of the toroid, 
and then decreases with the extension.
This behavior repeats many times and the final curve shows a sawtooth structure.
For the minor radius $r_0$, we found that it is a downstairs function, 
decreasing simultaneously with $W$.  
When a loop is pulled out of the toroid, the number of the chain strands 
in the annular tube of the toroid decreases by one, 
and thus $r_0$ exhibits a discontinuous jump.
These findings show that at the first stage of stretching, 
the toroid decreases its size constantly to some value. 
The size then oscillates around this value ($R_0\simeq 5$ in this study) 
while the toroid starts to loose its loops, one by one, 
during the stretching process.  

\begin{figure}
\begin{center}
(a)\\
\includegraphics[angle=0,width=\figurewidth]{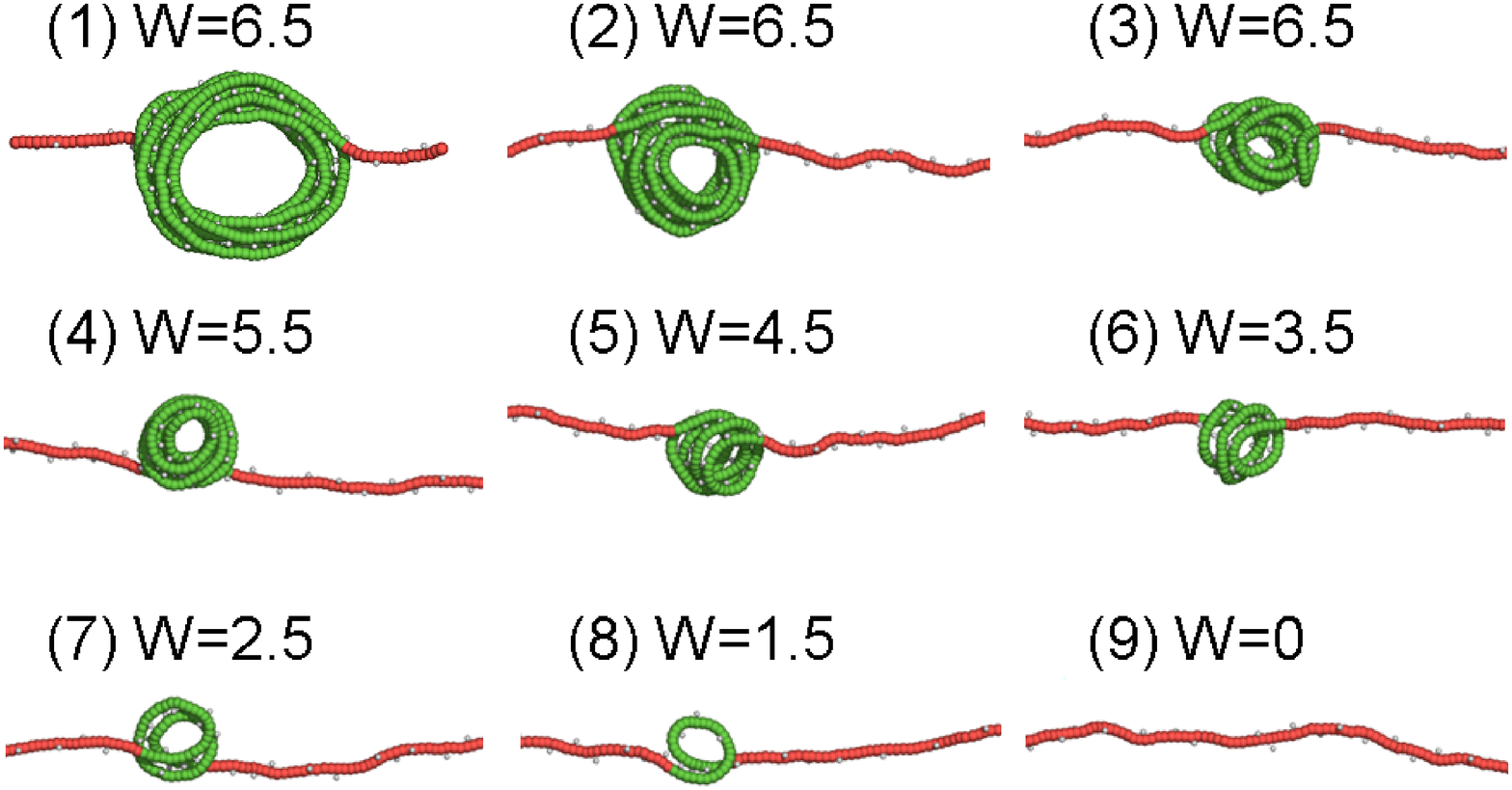}\\ 
(b)\\
\includegraphics[angle=270,width=\figurewidth]{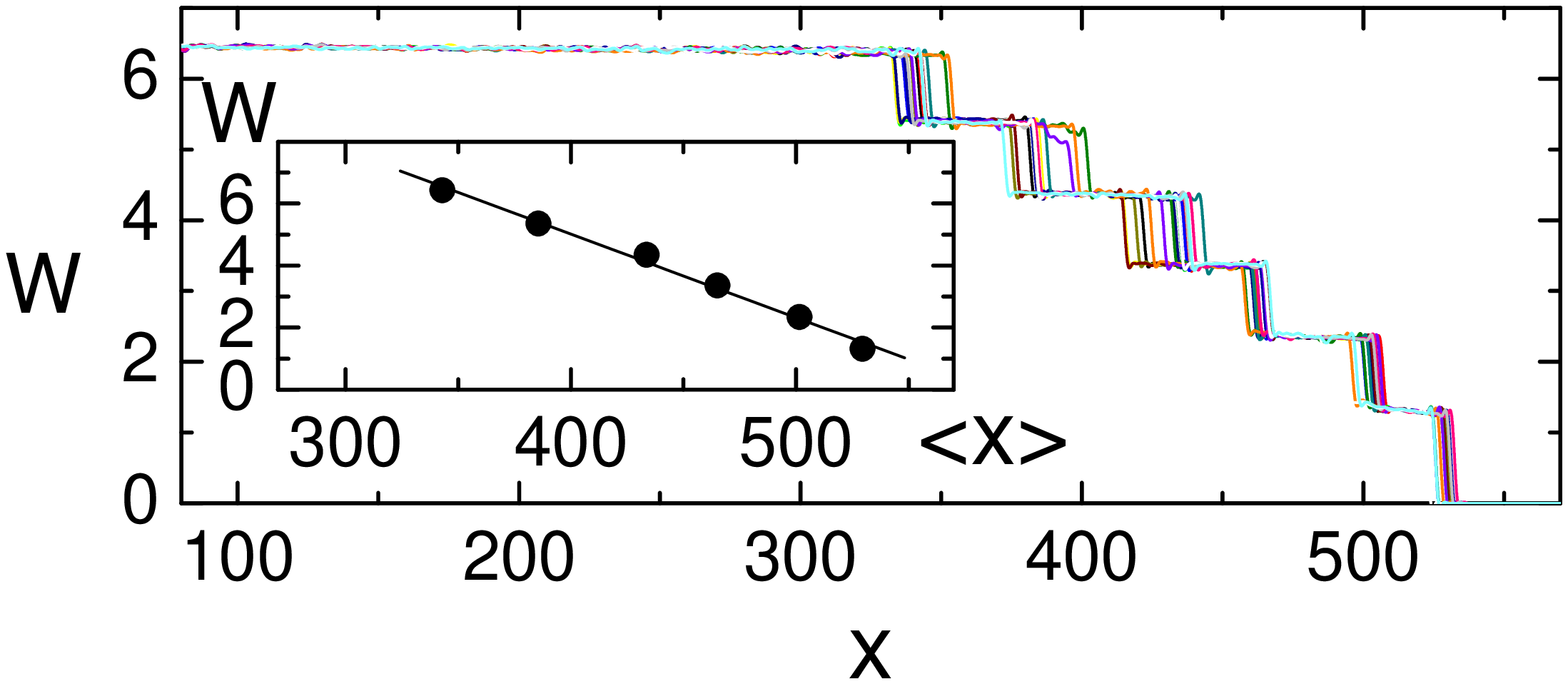}
\caption{ (a) Snapshots of simulations. The toroid part of the chain is
colored in light (green) color. 
(b) $W$ as a function of $x$ for the 20 independent runs. 
The inset shows the averaged position $\left< x\right>$ 
to occur a loss of one loop under stretching.}
\label{fig:snapshots}
\end{center}
\end{figure}

This structural transition has been approved by the 
snapshots, shown in Fig.~\ref{fig:snapshots}(a).
The snapshots clearly show that the diameter of the toroid firstly decreases to a critical 
value (pictures (1) to (3)) and then keeps the value around this value for
the rest of the loop releasing process (pictures (4) to (8)).  
The breaking of the toroid into several loops (or toroids) is not 
observed in the simulations. 
We have verified that the other independent runs 
show the consistent  elastic response.
The $W$ curves for these 20 independent runs are plotted
in Fig.~\ref{fig:snapshots}(b). 
The consistency of these curves confirms that the whole process 
is repeatable and the nonequilibrium effect  has been 
minimized by our slow pulling speed. 
Therefore, the results can be used to understand 
the elastic response in a near-equilibrium condition. 
We, furthermore, calculated the mean value of $x$ at which $W$ 
decreases by a step. The results, $W$ vs.~$\left<x\right>$, are plotted 
in the inset of Fig.~\ref{fig:snapshots}(b).
We see that $\left< x\right>$ lies on a straight line, 
which demonstrates that the circumference of a loop is a constant.
The slope of the line corresponds to a circumference $\Delta \left< x\right>=37.03$
per loop, or equivalently a circle of radius 5.9.
This radius is consistent with the value of $R_0$ at the peaks of
the sawteeth in Fig.~\ref{fig:R0_r0}. 

It is known that the persistence length of DNA is about 50nm.
Consequently, when DNA is collapsed into a toroid, 
the periphery length of the toroid is approximately 300nm. 
Murayama \textit{et al.}~\cite{Murayama03} have shown that 
the stick-release pattern has a periodicity characterized 
by a distance of 300nm. 
This result supports the loop-by-loop unfolding of DNA, 
as do our simulations. 
Moreover, they found that the higher the salt concentration,
the earlier the stick-release pattern happens in the stretching.
Thus, the chain looses its first loop at a smaller extention
position, according to our results. 
This can be understood by the phenomena of reentrant transition, 
in which a condensed DNA becomes less and less stable while 
the salt concentration is increased~\cite{Nguyen00,Wei07}.
Because the loss of the first loop of toroid 
occurs at a larger $R_0$ at higher salt concentration,
we predict that the periodicity of the stick-release 
pattern increases with salt concentration.

\begin{figure}
\begin{center}
\includegraphics[angle=270,width=\figurewidth]{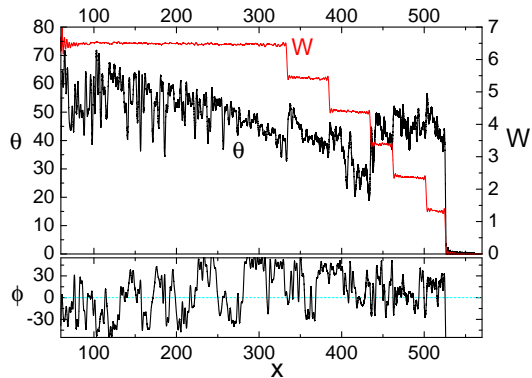} 
\caption{$\theta$ and $\phi$, both in unit of angle degree ($^{\circ}$), as a function of $x$.
$W$ is plotted to indicate the occurrence of the loop-by-loop unfolding.}
\label{fig:zenith_angle}
\end{center}
\end{figure}

Finally, we studied the zenith angle $\theta$, measured from the
pulling-force direction to the normal direction of the toroid surface, 
to give insight of the unwrapping pathway of a toroidal condensate under tension.   
$\theta$ is plotted in Fig.~\ref{fig:zenith_angle} as a function of $x$. 
We found that the chain was stretched from a starting zenith angle of about $65^{\circ}$.
In the stretching process, $\theta$ showed piecewise decreasing 
behavior with $x$, happened coherently with the change of the 
winding number. 
Therefore, the toroid normal was pulled, over again, toward 
the force direction during the stretching and 
swung back to its early direction at each moment when
the toroid released a loop. 
It is the sudden drop of the chain tension which  
restores the normal vector back to its early direction. 
We also calculated the azimuthal angle $\phi$ of the normal vector, 
plotted in Fig~\ref{fig:zenith_angle} too, which shows that 
the projection of the vector on the plane perpendicular to the pulling direction
fluctuated around a reference direction.
Kuli{\'c} and Schiessel predicted a \textit{flipping-up-and-down} pathway of transition 
for the normal vector of a spool of DNA helically coiled onto a histone 
protein~\cite{Kulic04}. 
Such transition was not observed in this study. 
This finding suggests an important role played by the histone protein.
For a DNA-histone spool under tension, the radius of the spool is restricted 
by the histone core. To pull a loop out of the spool inevitably induces a 
flipping-up-and-down movement of the spool.
On the other hand, for a spool of DNA wrapped onto itself by multivalent counterions,
a loop can be released gradually by decreasing the radius; 
therefore, the motion of the spool is more gentle and $\theta$ oscillates in a small
range of angle, as seen in our case.
This delicate difference of the dynamics between these two systems  
will be clarified more clearly in the future.

In summary, we have demonstrated that the loop-by-loop unfolding of
a toroidal PE produces the stick-release pattern in FEC. 
We have shown how the internal structure of a toroidal condensate 
and its normal vector change during stretching.
The results give deep insight of the elastic response and the structural 
transition of condensed DNA molecules being stretched.   

\begin{acknowledgments}
This work is supported by the National Science Council (Grant No.~NSC 97-2112-M-007-007-MY3).
Computing resources are supported by the National Center for High-performance Computing. 
P.~C.-L.~acknowledges the graduate fellowship from the Taiwan Semiconductor Manufacturing Company. 
\end{acknowledgments}

%
%
%


\begin{thebibliography}{28}
\expandafter\ifx\csname natexlab\endcsname\relax\def\natexlab#1{#1}\fi
\expandafter\ifx\csname bibnamefont\endcsname\relax
  \def\bibnamefont#1{#1}\fi
\expandafter\ifx\csname bibfnamefont\endcsname\relax
  \def\bibfnamefont#1{#1}\fi
\expandafter\ifx\csname citenamefont\endcsname\relax
  \def\citenamefont#1{#1}\fi
\expandafter\ifx\csname url\endcsname\relax
  \def\url#1{\texttt{#1}}\fi
\expandafter\ifx\csname urlprefix\endcsname\relax\def\urlprefix{URL }\fi
\providecommand{\bibinfo}[2]{#2}
\providecommand{\eprint}[2][]{\url{#2}}


\bibitem[{\citenamefont{Vijayanathan et~al.}(2002)\citenamefont{authors}}]{Vijayanathan02}
  \bibinfo{author}{\bibfnamefont{V.}~\bibnamefont{Vijayanathan}},
  \bibinfo{author}{\bibfnamefont{T.}~\bibnamefont{Thomas}}, \bibnamefont{and}
  \bibinfo{author}{\bibfnamefont{T. J.} \bibnamefont{Thomas}},
  \bibinfo{journal}{Biochemistry} \textbf{\bibinfo{volume}{41}},
  \bibinfo{pages}{14085} (\bibinfo{year}{2002}).

\bibitem[{\citenamefont{Smith et~al.}(1992)\citenamefont{Smith, Finzi, and Bustamante.}}]{Smith92}
  \bibinfo{author}{\bibfnamefont{S. B.}~\bibnamefont{Smith}},
  \bibinfo{author}{\bibfnamefont{L.}~\bibnamefont{Finzi}},  \bibnamefont{and}
  \bibinfo{author}{\bibfnamefont{C.}~\bibnamefont{Bustamante}},
  \bibinfo{journal}{Science} \textbf{\bibinfo{volume}{258}},
  \bibinfo{pages}{1122} (\bibinfo{year}{1992});
  \bibinfo{author}{\bibfnamefont{E.-L.}~\bibnamefont{Florin}},
  \bibinfo{author}{\bibfnamefont{V. T.}~\bibnamefont{Moy}},  \bibnamefont{and}
  \bibinfo{author}{\bibfnamefont{H. E.}~\bibnamefont{Gaub}},
  \bibinfo{journal}{Science} \textbf{\bibinfo{volume}{264}},
  \bibinfo{pages}{415} (\bibinfo{year}{1994});
  \bibinfo{author}{\bibfnamefont{R. M.}~\bibnamefont{Simmons}},
  \bibinfo{author}{\bibfnamefont{J. T.}~\bibnamefont{Finer}},
  \bibinfo{author}{\bibfnamefont{S.}~\bibnamefont{Chu}},  \bibnamefont{and}
  \bibinfo{author}{\bibfnamefont{J. A.}~\bibnamefont{Spudich}},
  \bibinfo{journal}{Biophys J.} \textbf{\bibinfo{volume}{70}},
  \bibinfo{pages}{1813} (\bibinfo{year}{1996});
  \bibinfo{author}{\bibfnamefont{C.}~\bibnamefont{Bustamante}},
  \bibinfo{author}{\bibfnamefont{Z.}~\bibnamefont{Bryant}}, \bibnamefont{and}
  \bibinfo{author}{\bibfnamefont{S. B.}~\bibnamefont{Smith}},
  \bibinfo{journal}{Nature} \textbf{\bibinfo{volume}{421}},
  \bibinfo{pages}{423} (\bibinfo{year}{2003}).

\bibitem[{\citenamefont{Baumann et~al.}(2000)\citenamefont{Baumann, Bloomfield, Smith, Bustamante, Wang, and Block.}}]{Baumann00}
  \bibinfo{author}{\bibfnamefont{C. G.}~\bibnamefont{Baumann}},
  \bibinfo{author}{\bibfnamefont{V. A.}~\bibnamefont{Bloomfield}},
  \bibinfo{author}{\bibfnamefont{S. B.}~\bibnamefont{Smith}},
  \bibinfo{author}{\bibfnamefont{C.}~\bibnamefont{Bustamante}},
  \bibinfo{author}{\bibfnamefont{M. D.}~\bibnamefont{Wang}}, \bibnamefont{and}
  \bibinfo{author}{\bibfnamefont{S. M.}~\bibnamefont{Block}},
  \bibinfo{journal}{Biophys. J.} \textbf{\bibinfo{volume}{78}},
  \bibinfo{pages}{1965} (\bibinfo{year}{2000}).

\bibitem[{\citenamefont{Murayama et~al.}(2001)\citenamefont{Murayama and Sano.}}]{Murayama01}
  \bibinfo{author}{\bibfnamefont{Y.}~\bibnamefont{Murayama}} \bibnamefont{and}
  \bibinfo{author}{\bibfnamefont{M.}~\bibnamefont{Sano}}, 
  \bibinfo{journal}{J. Phys. Soc. Jpn.} \textbf{\bibinfo{volume}{70}},
  \bibinfo{pages}{345} (\bibinfo{year}{2001}).

\bibitem[{\citenamefont{Murayama et~al.}(2003)\citenamefont{Murayama Sakamaki, and Sano.}}]{Murayama03}
  \bibinfo{author}{\bibfnamefont{Y.}~\bibnamefont{Murayama}}, 
  \bibinfo{author}{\bibfnamefont{Y.}~\bibnamefont{Sakamaki}}, \bibnamefont{and}
  \bibinfo{author}{\bibfnamefont{M.}~\bibnamefont{Sano}},
  \bibinfo{journal}{Phys. Rev. Lett.} \textbf{\bibinfo{volume}{90}},
  \bibinfo{pages}{018102} (\bibinfo{year}{2003}).

\bibitem[{\citenamefont{Marko et~al.}(1995)\citenamefont{Marko and Siggia}}]{Marko95}
  \bibinfo{author}{\bibfnamefont{J. F.}~\bibnamefont{Marko}} \bibnamefont{and} 
  \bibinfo{author}{\bibfnamefont{E. D.}~\bibnamefont{Siggia}},
  \bibinfo{journal}{Macromolecules} \textbf{\bibinfo{volume}{28}},
  \bibinfo{pages}{8759} (\bibinfo{year}{1995}).

\bibitem[{\citenamefont{Bloomfield}(1995)\citenamefont{Bloomfield}}]{Bloomfield96}
\bibinfo{author}{\bibfnamefont{V.~A.} \bibnamefont{Bloomfield}},
  \bibinfo{journal}{Curr. Opin. Struct. Biol.} \textbf{\bibinfo{volume}{6}},
  \bibinfo{pages}{334} (\bibinfo{year}{1996});
  \bibinfo{author}{\bibfnamefont{N. V.}~\bibnamefont{Hud}}, 
  \bibinfo{author}{\bibfnamefont{K. H.}~\bibnamefont{Downing}}, \bibnamefont{and}
  \bibinfo{author}{\bibfnamefont{R.}~\bibnamefont{Balhorn}},
  \bibinfo{journal}{Proc. Natl. Acad. Sci. USA} \textbf{\bibinfo{volume}{92}},
  \bibinfo{pages}{3581} (\bibinfo{year}{1995}).

\bibitem[{\citenamefont{Nguyen et~al.}(2000)\citenamefont{Nguyen, Rouzina, and
  Shklovskii}}]{Nguyen00}
  \bibinfo{author}{\bibfnamefont{T.~T.} \bibnamefont{Nguyen}},
  \bibinfo{author}{\bibfnamefont{I.}~\bibnamefont{Rouzina}}, \bibnamefont{and}
  \bibinfo{author}{\bibfnamefont{B.~I.} \bibnamefont{Shklovskii}},
  \bibinfo{journal}{J. Chem. Phys.} \textbf{\bibinfo{volume}{112}},
  \bibinfo{pages}{2562} (\bibinfo{year}{2000}).

\bibitem[{\citenamefont{Wada et~al.}(2002)\citenamefont{Wada, Murayama, and Sano.}}]{Wada02}
  \bibinfo{author}{\bibfnamefont{H.}~\bibnamefont{Wada}}, 
  \bibinfo{author}{\bibfnamefont{Y.}~\bibnamefont{Murayama}}, \bibnamefont{and}
  \bibinfo{author}{\bibfnamefont{M.}~\bibnamefont{Sano}},
  \bibinfo{journal}{Phys. Rev. E} \textbf{\bibinfo{volume}{66}},
  \bibinfo{pages}{061912} (\bibinfo{year}{2002}).

\bibitem[{\citenamefont{Ueda et~al.}(1996)\citenamefont{Ueda and Yoshikawa.}}]{Ueda96}
  \bibinfo{author}{\bibfnamefont{M.}~\bibnamefont{Ueda}} \bibnamefont{and}
  \bibinfo{author}{\bibfnamefont{K.}~\bibnamefont{Yoshikawa}}, 
  \bibinfo{journal}{Phys. Rev. Lett.} \textbf{\bibinfo{volume}{77}},
  \bibinfo{pages}{2133} (\bibinfo{year}{1996}).

\bibitem[{\citenamefont{Wada et~al.}(2005)\citenamefont{Wada, Murayama, and Sano.}}]{Wada05}
  \bibinfo{author}{\bibfnamefont{H.}~\bibnamefont{Wada}}, 
  \bibinfo{author}{\bibfnamefont{Y.}~\bibnamefont{Murayama}}, \bibnamefont{and}
  \bibinfo{author}{\bibfnamefont{M.}~\bibnamefont{Sano}},
  \bibinfo{journal}{Phys. Rev. E} \textbf{\bibinfo{volume}{72}},
  \bibinfo{pages}{041803} (\bibinfo{year}{2005}).

\bibitem[{\citenamefont{Kulic et~al.}(2004)\citenamefont{eyc}}]{Kulic04}
  \bibinfo{author}{\bibfnamefont{I. M.}~\bibnamefont{Kuli{\'c}}} \bibnamefont{and}
  \bibinfo{author}{\bibfnamefont{H.}~\bibnamefont{Schiessel}},
  \bibinfo{journal}{Phys. Rev. Lett} \textbf{\bibinfo{volume}{92}},
  \bibinfo{pages}{228101} (\bibinfo{year}{2004}).

\bibitem[{\citenamefont{Marenduzzo et~al.}(2004)\citenamefont{Marenduzzo, Martian, and Rose}}]{Marenduzzo04}
  \bibinfo{author}{\bibfnamefont{D.}~\bibnamefont{Marenduzzo}}, 
  \bibinfo{author}{\bibfnamefont{A.}~\bibnamefont{Martian}}, 
  \bibinfo{author}{\bibfnamefont{A.}~\bibnamefont{Rosa}}, \bibnamefont{and}
  \bibinfo{author}{\bibfnamefont{A.}~\bibnamefont{Seno}},
  \bibinfo{journal}{Eur. Phys. J. E} \textbf{\bibinfo{volume}{15}},
  \bibinfo{pages}{83} (\bibinfo{year}{2004}).

\bibitem[{\citenamefont{Lee et~al.}(2005)\citenamefont{authors}}]{Lee04}
  \bibinfo{author}{\bibfnamefont{N.-K.}~\bibnamefont{Lee}} \bibnamefont{and}
  \bibinfo{author}{\bibfnamefont{D.}~\bibnamefont{Thirumalai}}
  \bibinfo{journal}{Biophys. J.} \textbf{\bibinfo{volume}{86}},
  \bibinfo{pages}{2641} (\bibinfo{year}{2004}).

\bibitem[{\citenamefont{Khan et~al.}(2005)\citenamefont{Khan and Chan}}]{Khan05}
  \bibinfo{author}{\bibfnamefont{M. O.}~\bibnamefont{Khan}} \bibnamefont{and}
  \bibinfo{author}{\bibfnamefont{D. Y. C.}~\bibnamefont{Chan}}
  \bibinfo{journal}{Macromolecules} \textbf{\bibinfo{volume}{38}},
  \bibinfo{pages}{3017} (\bibinfo{year}{2005}).

\bibitem[{not({\natexlab{b}})}]{note-tools}
 \bibinfo{note}{The simulations were run using a modified LAMMPS package.
  Refer to http://lammps.sandia.gov/ for LAMMPS.}

\bibitem[{\citenamefont{Stevens}(2001)\citenamefont{Stevens}}]{Stevens01}
  \bibinfo{author}{\bibfnamefont{M. J.}~\bibnamefont{Stevens}}, 
  \bibinfo{journal}{Biophys. J.} \textbf{\bibinfo{volume}{80}},
  \bibinfo{pages}{130} (\bibinfo{year}{2001}).

\bibitem[{\citenamefont{Wei et~al.}(2007)\citenamefont{Wei and Hsiao}}]{Wei07}
  \bibinfo{author}{\bibfnamefont{Y.-F.}~\bibnamefont{Wei}} \bibnamefont{and} 
  \bibinfo{author}{\bibfnamefont{P.-Y.}~\bibnamefont{Hsiao}}, 
  \bibinfo{journal}{J. Chem. Phys.} \textbf{\bibinfo{volume}{127}},
  \bibinfo{pages}{064901} (\bibinfo{year}{2007}).

\bibitem[{\citenamefont{Tamashiro et~al.}(2000)\citenamefont{Tamashiro and Schiessel}}]{Tamashiro00}
  \bibinfo{author}{\bibfnamefont{M. N.}~\bibnamefont{Tamashiro}} \bibnamefont{and}
  \bibinfo{author}{\bibfnamefont{H.}~\bibnamefont{Schiessel}},
  \bibinfo{journal}{Macromolecules.} \textbf{\bibinfo{volume}{33}},
  \bibinfo{pages}{5263} (\bibinfo{year}{2000}).

\bibitem[{\citenamefont{Odijk}(1995)\citenamefont{authors}}]{Odijk95}
  \bibinfo{author}{\bibfnamefont{T.}~\bibnamefont{Odijk}}, 
  \bibinfo{journal}{Macromolecules} \textbf{\bibinfo{volume}{28}},
  \bibinfo{pages}{7016} (\bibinfo{year}{1995});
  \bibinfo{author}{\bibfnamefont{M. D.}~\bibnamefont{Wang}}, 
  \bibinfo{author}{\bibfnamefont{H.}~\bibnamefont{Yin}}, 
  \bibinfo{author}{\bibfnamefont{R.}~\bibnamefont{Landick}}, 
  \bibinfo{author}{\bibfnamefont{J.}~\bibnamefont{Gelles}}, \bibnamefont{and} 
  \bibinfo{author}{\bibfnamefont{S. M.}~\bibnamefont{Block}},
  \bibinfo{journal}{Biophys. J.} \textbf{\bibinfo{volume}{72}},
  \bibinfo{pages}{1335} (\bibinfo{year}{1997}).

\bibitem[{\citenamefont{Hsiao}(2006)\citenamefont{Hsiao}}]{Hsiao06b}
  \bibinfo{author}{\bibfnamefont{P.-Y.}~\bibnamefont{Hsiao}}, 
  \bibinfo{journal}{Macromoleucles} \textbf{\bibinfo{volume}{39}},
  \bibinfo{pages}{7125} (\bibinfo{year}{2006}).

\end{thebibliography}
\end{document}